%% file: main.tex
\documentclass[conference]{IEEEtran}

\usepackage{cite}
\usepackage{listings}
\usepackage{subcaption}
\input{format}
\begin{document}

\title{A Requirements-Driven Platform for Validating Field Operations of Small Uncrewed Aerial Vehicles}

\author{

\IEEEauthorblockN{Ankit Agrawal}
\IEEEauthorblockN{Bohan Zhang}
\IEEEauthorblockN{Yashaswini Shivalingaiah}
\IEEEauthorblockA{\textit{Department of Computer Science} \\
\textit{Saint Louis University}\\
St Louis, MO, USA \\
ankit.agrawal.1@slu.edu}
\and

\IEEEauthorblockN{Michael Vierhauser}
\IEEEauthorblockA{\textit{LIT Secure and Correct Systems Lab} \\
\textit{Johannes Kepler University Linz}\\
Linz, Austria \\
michael.vierhauser@jku.at}
\and
\IEEEauthorblockN{Jane Cleland-Huang}
\IEEEauthorblockA{\textit{Computer Science And Engineering} \\
\textit{University Of Notre Dame}\\
Notre Dame, IN, USA \\
janehuang@nd.edu}
}


\maketitle

\begin{abstract}

Flight-time failures of small Uncrewed Aerial Systems (sUAS) can have a severe impact on people or the environment. Therefore, sUAS applications must be thoroughly evaluated and tested to ensure their adherence to specified requirements, and safe behavior under real-world conditions, such as poor weather, wireless interference, and satellite failure. However, current simulation environments for autonomous vehicles, including sUAS, provide limited support for validating their behavior in diverse environmental contexts and moreover, lack a test harness to facilitate structured testing based on system-level requirements. We address these shortcomings by eliciting and specifying requirements for an sUAS testing and simulation platform, and developing and deploying it. The constructed platform, \DWFull~(\DW), allows sUAS developers to define the operating context, configure multi-sUAS mission requirements, specify safety properties, and deploy their own custom sUAS applications in a high-fidelity 3D environment. The \DW Monitoring system collects runtime data from sUAS and the environment, analyzes compliance with safety properties, and captures violations. We report on two case studies in which we used our platform prior to real-world sUAS deployments, in order to evaluate sUAS mission behavior in various environmental contexts. Furthermore, we conducted a study with developers and found that \DW simplifies the process of specifying requirements-driven test scenarios and analyzing acceptance test results.

\end{abstract}
\vspace{0.5em}

\input{sec_01_introduction}

\input{sec_02_overview}
\input{sec_03_env_configuration}
\input{sec_04_architecture}
\input{sec_06_evaluation}

\input{sec_07_threats}

\input{sec_08_related_work}
\section{Conclusion and Future Work}
\label{tb_sec:conclusion}


In this paper, we have presented \DW, a platform that software engineers developing novel sUAS applications can use to specify diverse requirements for multi-sUAS missions, define acceptance tests, and deploy their own sUAS missions into a realistic simulation environment. We have validated the feasibility and usefulness of our platform through applying it to two different sUAS missions, and through evaluating sUAS developers' perception of \DW. Our platform supports requirements validation and black-box testing of entire missions, and provides critical support for deploying validated sUAS applications into the physical world. This paper contributes to the field of requirements engineering by offering a novel approach to specifying and testing complex sUAS missions in a realistic simulation environment. In future work, we plan to extend our platform to support specific scenarios, such as delivery or surveillance, and through integrating more advanced features such as fault injection and recovery mechanisms.


\section*{Data Availability}
\label{sec:data_availability}
We provide a list of sUAS incidents, evaluation data, web app for designing test scenarios, \DW simulation engine package, and user study materials in our public Github repository\footnote{\url{https://github.com/UAVLab-SLU/RE-23-Supp-Materials}}.  


\section*{Acknowledgment} The work described in this paper was primarily funded under NSF Grant 1931962 and partially funded by the  Linz Institute of Technology (LIT-2019-7-INC-316). 

\balance


\bibliographystyle{IEEEtran}
\bibliography{ICSE2023.bib}

\end{document}

%% file: format.tex
\usepackage{booktabs} 
\usepackage{color}
\usepackage{hyperref}
\usepackage{multirow}
\usepackage{pifont}
\usepackage{afterpage}
\usepackage{rotating}
\usepackage{array}
\usepackage[x11names,dvipsnames]{xcolor}
\usepackage[nointegrals]{wasysym}
\usepackage{amsmath}
\usepackage{txfonts}
\newcolumntype{x}[1]{>{\raggedright\arraybackslash}p{#1}}

\usepackage{makecell}
\usepackage{graphicx}
\usepackage[font=footnotesize]{caption}
\usepackage[font=footnotesize]{subcaption}
\usepackage[most]{tcolorbox}

\usepackage{multirow}
\usepackage{soul}
\setul{.3ex}{}

\usepackage{url}
\usepackage{lscape}
\usepackage{rotating}
\usepackage{tabularx}

\usepackage{colortbl}
\usepackage{ragged2e}
\usepackage{verbatim}
\usepackage{listings}
\usepackage{arydshln}
\usepackage[neverdecrease]{paralist}

\usepackage{flushend}
\usepackage{enumitem}
\usepackage{CJKutf8}
\usepackage{tikz}
\usepackage{tabulary}

\usepackage{wrapfig}
\usepackage{floatflt}

\usepackage{bigstrut}
\usepackage{threeparttablex}

\usepackage{amssymb}
\usepackage{diagbox}
\usepackage[ruled,vlined, linesnumbered]{algorithm2e}
\definecolor{grayhighlight}{rgb}{.8,0.8,0.8}
\definecolor{LightCyan}{rgb}{0.88,1,1}
\definecolor{LightAmber}{rgb}{0.98,0.89,0.74}
\definecolor{pink}{rgb}{0.9,0,0.9}
\definecolor{gray}{rgb}{0.95,0.95,0.85}
\definecolor{mauve}{rgb}{0.1,0.7,0.2}
\definecolor{bg}{rgb}{0.95,0.95,0.9}
\definecolor{dkgreen}{rgb}{0,0.6,0}
\definecolor{gray2}{rgb}{0.95,0.95,0.95}
\definecolor{gray3}{rgb}{0.20,0.20,0.20}
\definecolor{lightgray}{rgb}{0.6,0.6,0.6}

\usepackage{colortbl}

\usepackage{balance}
\newcolumntype{L}[1]{>{\raggedright\let\newline\\\arraybackslash}p{#1}} 
\newcolumntype{C}[1]{>{\centering\let\newline\\\arraybackslash}p{#1}} 
\newcolumntype{R}[1]{>{\raggedleft\let\newline\\\arraybackslash}p{#1}}

\usepackage[utf8]{inputenc}
\usepackage{epstopdf}

\usepackage{float}
\usepackage{xspace}

\DeclareRobustCommand{\hlcyan}[1]{{\sethlcolor{LightCyan1}\hl{#1}}}

\newcolumntype{M}[1]{>{\centering\arraybackslash}m{#1}}

\newcommand{\DWFull}{DroneReqValidator}
\newcommand{\DW}{DRV\xspace}

\newcommand{\jch}[1]{{\textcolor{purple}{#1}}}

\newcommand{\simpleitem}[1]{\noindent$-$ \emph{#1}}
\newcommand{\etal}{{\textit{et~al.}}\xspace}

\newtcolorbox[auto counter]{simreq}[1]{title={\bfseries Requirement},drop shadow={black!10!white},
  coltitle=white, colframe=white!25!black, boxsep=2pt,left=4pt,right=4pt,top=3pt,bottom=3pt,  sharpish corners}

  \providecommand\BibTeX{{%
    \normalfont B\kern-0.5em{\scshape i\kern-0.25em b}\kern-0.8em\TeX}}

\newcommand{\citesec}[1]{Section~\ref{sec:#1}}
\newcommand{\citefig}[1]{Figure~\ref{fig:#1}}

\newcommand{\citetable}[1]{{Table}~\ref{tab:#1}}

\lstdefinestyle{jsonlistingstyle}{
  language=json,
  numbers=left,
  stepnumber=1,
  numbersep=8pt,
  tabsize=2,
  showspaces=false,
  basicstyle=\footnotesize\ttfamily,
   numberstyle=\scriptsize,
    showstringspaces=false,
    breaklines=true,
    frame=lines,
    backgroundcolor=\color{background},
}

%% file: sec_01_introduction.tex
\section{Introduction}
With the rise of artificial intelligence, small Uncrewed Aerial Systems (sUAS) are imbued with increasingly complex decision-making capabilities, in order to perform missions autonomously in diverse environmental conditions~\cite{ilachinski2017artificial}. 
As failures during operation can lead to severe accidents that are harmful to people, physical structures, or the environment, it is essential to specify safety requirements, design effective solutions, and establish a robust testing process, infrastructure, and corresponding monitoring tools for validating that the system satisfies its requirements prior to deployment ~\cite{castellanos2019modular,zhang2016understanding,ali2015u,antkiewicz2020modes}. Environmental conditions, and their diverse combinations, especially those at the boundaries of an sUAS' operating capacity,  can impact the behavior of an sUAS in unpredictable ways, and therefore, many accounts of sUAS flight failures due to problems such as radio interference~\cite{nasa-aviation}, or high winds~\cite{CONFIG-W1,CONFIG-W2} have occurred. This, in turn, means that functional tests must be executed under diverse conditions. For example, the requirement that \emph{``An sUAS shall complete a flight composed of multiple waypoints in wind gusts of 23mph without colliding with stationary objects, the terrain, or other aircraft''} needs to be operationalized within diverse test scenarios that specify the specific flight details, as well as additional environmental attributes such as  wind direction, temperature, precipitation, visibility, and geographical information.

Performing rigorous software verification and validation (V\&V) on Cyber-Physical Systems (CPS) in general, and sUAS in particular, is a time-consuming process that typically involves a combination of simulations and real-world testing to validate the correctness of system behavior under a range of conditions~\cite{anda2019arithmetic,jung2009real,zheng2015perceptions}. Furthermore, many tests cannot easily be conducted on physical sUAS, especially those that target, or even exceed operational boundaries, such as flying in extreme weather conditions or in (too) close proximity to objects or humans. However, critical differences between the simulation and the real-world environment can result in substantial back-and-forth testing between physical testing sites and developers, extending project development times and increasing costs. This problem is primarily attributable to (a)~lack of tool support for developing realistic scenario simulations, (b)~difficulties in identifying and/or modeling edge-case scenarios in the real-world environment, (c)~isolated simulation environments that fail to consider interactions with sensors and physical devices used by humans to interact with the system, and (d)~the lack of a structured process and platform for specifying, executing, analyzing, and testing diverse system requirements. 

In practice, for the domain of sUAS, developers currently rely on simulations using 2D maps~\cite{dronology,fernando2013modelling,rodriguez2015design} or 3D simulation environments, such as Gazebo~\cite{gazebo} or AirSim~\cite{shah2018airsim}. 
Gazebo~\cite{koenig2004design}, for example, facilitates sUAS simulations with limited automated support for incorporating realistic landscapes \cite{abbyasov2020automatic} and weather conditions, while AirSim provides high-fidelity weather simulations, but it lacks realistic flight conditions such as simulating real-world airspace restrictions, and mission-specific environmental elements, such as simulating a drowning person in a river to support search-and-rescue test scenarios~\cite{shah2018airsim, madaan_airsim_2020}. These existing simulation environments rely more upon an ad-hoc, trial-and-error testing approach with limited support for specifying real-world test scenarios~\cite{zheng2015perceptions}, and provide even less support for a requirements-driven test environment in which diverse scenarios are generated and executed for given requirements.

In this paper, we address this challenge by presenting a new platform, \DWFull, referred to as \DW in the remainder of the paper, for supporting the creation of requirements-driven test scenarios. We employed design science~\cite{wieringa2014design} to collect and specify clear design and development objectives of \DW (discussed in Section \ref{sec:overview}), based on identified key challenges. Specifically during simulation testing, and related to simulating volatile \emph{high-fidelity environmental conditions} that affect sUAS behavior (discussed in Section~\ref{sec:configproperties}). Our platform allows developers or testers to specify environmental conditions, configure sUAS sensor capabilities, and specify test properties to validate system-level requirements. Based on specifications, the platform generates the simulation environment and deploys sUAS with configured sensor information. \DW automatically monitors the specified test properties for violation and generates an acceptance test report containing detailed simulation analytics. By using \DW, developers can investigate the capabilities and limitations of their sUAS applications against system-level requirements prior to field deployment.

\noindent The contributions of this paper are therefore as follows:\\\vspace{-15pt}
\begin{enumerate}[leftmargin=1.5em]
    \setlength\itemsep{+.1em}
    \item[(1)] We analyze real-world sUAS incidents to identify common points of failure and subsequently specify requirements for the \DW validation platform. Our aim is to enable sUAS developers to effectively validate critical mission behaviors within realistic contexts.
    \item[(2)] We describe the  set of safety-related, first-order runtime properties used by \DW to validate requirements for correct flight operations of sUAS.
    \item[(3)] We derive a structured end-to-end process for specifying various elements of test scenarios, for performing multi-sUAS fuzz testing activities, analyzing simulation results, and  evaluating \DW's fidelity for physical sUAS systems under test. This process advances the state-of-the-art in requirements-driven simulation platforms for autonomous vehicles.
    \end{enumerate}

 To evaluate our platform, we conducted a case study with two real-world cases in which \DW was used to validate requirements for our own drone system  composed of multiple autonomous and collaborating sUAS. DroneResponse \cite{nafeeconfiguring2022} assumed the role of the {\bf D}rone {\bf S}ystem {\bf u}nder {\bf T}est (DSuT).  In addition, we performed a preliminary study to evaluate sUAS developers' opinion of \DW and its ability for designing requirements-based tests.

The remainder of this paper is organized as follows. \citesec{overview} provides an overview of our \DW platform, and in \citesec{configproperties}, we further describe the set of features vital for an sUAS testing platform and describe \DW's ability for configuring the environment and runtime properties.
In \citesec{platform}, we comprehensively describe our platform architecture and the process of specifying and executing high-fidelity realistic environmental conditions and respective mission-specific tests are created 
We outline our evaluation set-up in \citesec{sim_eval}, and present our findings in Sections~\ref{sec:rq1} and \ref{sec:rq2}. We, finally, discuss threats to validity and related work in Sections~\ref{sec:threats} and \ref{sec:relwork}.

\begin{figure}[b!]
    \centering
    \vspace{-15pt}
    \includegraphics[width=0.90\columnwidth]{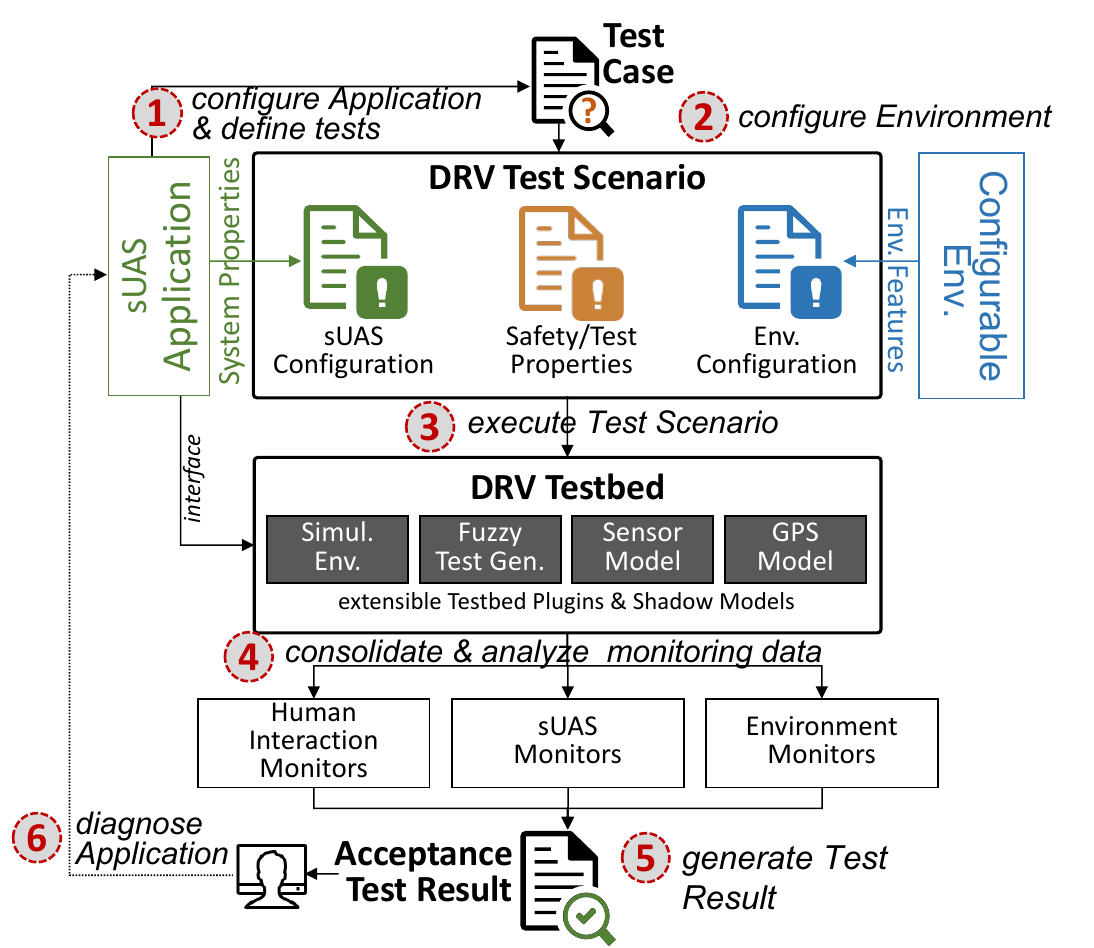}
    \caption{An overview of the \DW platform, showing how a user specifies real-world test scenarios for their own sUAS applications, executes them in diverse environmental conditions and evaluates their outcomes.}
    \label{fig:overview}
\end{figure}

%% file: sec_02_overview.tex
\section{Overview of \DW}
\label{sec:overview}

Testing and validating a CPS against its requirements requires more than ``just'' simulating its behavior in a virtual space, and involves systematic requirements elicitation, hardware and software testing including the definition and analysis of safety properties, and integration of the user and human interactions~\cite{ayerdi2020towards,abbaspour2015survey, zhou2018review,zheng2015perceptions}. Therefore, our primary objective of developing \DW is twofold.
\begin{itemize}
    \item \textbf{DO1}: Develop a simulation approach that goes beyond providing simplistic pass/fail test results.

    \item \textbf{DO2}: Automatically execute diverse test scenarios under realistic 3D environments to effectively detect safety-related issues in sUAS applications.
\end{itemize}

To accomplish these primary development objectives, we identified several features to incorporate in \DW. First, to validate a set of requirements, developers need the ability to \emph{specify, execute, and validate complex test scenarios} that include (1) realistic scenes with buildings, trees, and other landmarks, (2) interactive environments that include, for example, roads, people, fires, and traffic accidents, and (3) environmental factors such as weather, wireless interference, and satellite availability. 
Second, the environment needs a diverse \emph{selection of high-fidelity sUAS with sensors and actuators} such as controllable cameras and other sensors; all of which provide APIs for interfacing with the test applications. For the simulation to support diverse sUAS applications, it needs to allow a tester to deploy their own sUAS applications within the test environment with minimal effort. Third, the \emph{test platform needs to facilitate runtime monitoring}\cite{vierhauser2016reminds}, by providing monitors that enable users to specify properties and constraints and to subsequently monitor the runtime behavior of individual sUAS, their interactions with each other, with human actors involved in a mission, and their environment. Finally, bringing all of this together, the environment must provide users with the \emph{means to specify test scenarios for specific requirements, and enact those tests in the defined environment} whilst simultaneously selecting relevant constraints to be checked during the running simulation. 
We have designed \DW to address these requirements, and provide a high-level overview in~\citefig{overview}.

In Step 1, \DW users develop their own single- or multi-sUAS applications using supported flight controllers (with current support for PX4 \cite{meier2015px4} and Ardupilot \cite{px4,ardupilot}), and then specify test cases for validating requirements, based on real-world scenarios such as multi-sUAS area search. 

In Step 2, the users configure the environmental conditions (e.g., weather, signal, geographical regions) in which they need to execute their tests as per system-level requirements as well as safety properties (e.g., ``minimum horizontal and lateral separation distances between sUAS'') that sUAS are expected to maintain during the mission. The combination of the sUAS application, prescribed mission, and environment and safety properties specifications constitute a {\it Test Scenario}. 

In Step 3, the \DW simulation engine creates realistic environmental conditions including weather conditions, and realistic terrains and landscapes, deploys the desired number of sUAS in the environment, and configures sensor models of each sUAS as per requirements. When activated, the fuzzy test generator component of the platform generates multiple test scenarios by fuzzing the user-provided environmental configuration within the given range of values. The objective of this component is to examine the robustness of the sUAS application by analyzing the extent to which the system is able to perform as expected in adverse environmental conditions. Finally, after generating test cases, the platform simulates both the primary configured test and its fuzzy versions.

In Step 4, data is collected from the sUAS and the environment throughout the mission, and the \DW monitoring system continually analyzes the data for violations. Finally, in Step 5, \DW produces an analytics report that contains analysis of simulation results from each test case, comparisons across all fuzzy test cases, and a list of detected violations.

The {\bf key novelty of \DW} is that it allows developers to easily specify test scenarios in combination with environmental conditions, sUAS capabilities, and a set of monitorable properties indicative of system-level requirements satisfaction conditions. Developers can deploy and validate their own sUAS applications  under the specified test scenario using \DW prior to deployment in the physical world. Furthermore, \DW supports fuzzy testing~\cite{nguyen2022bedivfuzz,sun2022lawbreaker} of system-level requirements, allowing developers to identify the operating boundaries of their application under adverse environmental conditions. Additionally, \DW consolidates and analyzes runtime information from diverse sources in order to provide insights into passed and failed test properties for further analysis and error diagnosis (Step 6). 
In the following sections, we describe the configuration properties, platform, and architecture in more detail.

%% file: sec_03_env_configuration.tex
\section{Environmental Configuration}
\label{sec:configproperties}

\input{tables/tab_01_datasources}

As input to the overall \DW design process, we sought to identify a common set of configurable environmental factors relevant for validating sUAS applications. We used a  deductive (top-down) approach based on a set of well-known sUAS problems related to environmental terrain~\cite{Popovic2020}, GPS denial~\cite{kang2018damage} signal interference~\cite{DBLP:journals/jcm/DuangsuwanM21}, weather and lighting~\cite{DBLP:conf/icse-wain/AbrahamCBVAISC21}, human-operations~\cite{vierhauser2021hazard}, and sUAS physical failures~\cite{DBLP:journals/corr/abs-2207-08857}. To identify detailed information about each of these categories, including specific types of failures and potential ways of monitoring them, we further conducted a search for sUAS incident reports published by news services and regulatory bodies and reviewed scientific publications related to sUAS safety requirements. As part of our incident investigation, we searched for scientific publications on Google Scholar and Web of Science for ``sUAS/UAV safety, accidents, incidents'' and examined sUAS incident reports as depicted in Table~\ref{tab:incident_sources}.


\input{tables/tab_02_factors}

\subsection{Configuration Requirements}
The literature and incident analysis identified numerous ways in which environmental factors played a key role in sUAS incidents. We categorize them into six groups, provide examples for each, and summarize the key factors.

\simpleitem{Geographical Locations and Terrain:} sUAS are deployed on diverse missions across vastly different types of terrain including open farmland, forests, urban areas with tall buildings, and mountainous terrain. Other areas include protected airspace (e.g., in close proximity to airports or over national parks, and prisons). Reports document diverse incidents including an sUAS collision with a Hot Air Balloon over Boise, Idaho whilst flying without authorization in controlled airspace~\cite{Incident-Balloon}, a goose hit in Sweden~\cite{CONFIG-L4}, a construction crane in Kent, UK~\cite{CONFIG-L2}, and a mountain in Colorado~\cite{CONFIG-L3}. The simulation environment, therefore, needs to support diverse scenarios including stationary and moving objects, diverse types of terrain, and restricted airspace to allow test cases that validate whether an sUAS can complete its mission successfully whilst complying with all legal airspace regulations. Configurable properties, therefore, include \texttt{Region}; \texttt{Controlled airspace} (regulated airspace and no-fly-zone automatically retrieved from service providers, with the ability to upload additional regions of prohibited or constrained airspace); and \texttt{Objects} (stationary objects, e.g., cranes, and movable objects such as vehicles).

\simpleitem{Signal Loss and sUAS Communication:} Loss of communication between the sUAS and the remote pilot (e.g.,~\cite{CONFIG-S2}) can occur if the sUAS exceeds its range capabilities, is obstructed by an obstacle, or when electromagnetic interference otherwise disrupts the data link~\cite{CONFIG-S3}. It can impact telemetry between the sUAS and a handheld radio controller, or software-based Ground Control System (GCS) (e.g., MissionPlanner, QGroundControl~\cite{qground,mplanner}), or other forms of communication (e.g., WiFi or Mesh Radio) between ground-based software systems and onboard software applications which are wired to the onboard flight controller. Loss-of-signal alone does not cause a crash, as most sUAS automatically enter failsafe modes, such as return-to-launch (RTL), when communication is disrupted; however, as airspace becomes increasingly crowded, simple RTL commands could themselves cause incidents~\cite{Incident-RTLOverride}. Configurable \texttt{radio interference models}, therefore, need to be integrated into \DW to allow users to test sUAS response to unexpected loss of data link. 

\simpleitem{GPS Deprivation:} Most sUAS rely upon GPS for geolocation purposes, and as a result, loss of reliable GPS leads to sUAS crashes~\cite{CONFIG-S1}. Geolocation accuracy typically increases with the number of satellites, returning up to 2 meters of accuracy with 15 or more satellite connections, but decreasing rapidly as the number of connections decreases. sUAS systems can compensate for geolocation uncertainty, for example, by maintaining greater distances from terrain, buildings, or other sUAS, or by using sensors to help prevent collisions. The \DW environment, therefore, needs configurable \texttt{GPS models} that are able to predict GPS accuracy and/or inject loss of satellite faults into the test environment. 

\simpleitem{Weather and Lighting Factors:} Weather conditions greatly impact the functioning of both the sUAS' perception of the environment and its control algorithms~\cite{gao2021weather}. Wind speed is often cited as a cause of sUAS crashes, because high wind speed in a particular direction can negatively affect the ability of the sUAS to maintain its desired flight path~\cite{lundby2019towards}. Furthermore, rain, fog, and snow, as well as low lighting conditions in the environment, impair the ability of computer vision models to interpret the environment correctly~\cite{hamzeh2022improving}. Turbulent weather has resulted in several incidents, including failure to hold position \cite{CONFIG-W1}, or dislodging payload~\cite{CONFIG-W2}. Therefore, \DW must support configuration and simulation of various \texttt{weather conditions}, including wind direction, speeds, and gusts at different altitudes, precipitation types and levels, \texttt{lighting conditions}, and \texttt{visibility}.

\simpleitem{Human Interactions:} Numerous sUAS accidents can be accounted to human-related errors, caused by recklessness, lack of training, or poor user interface design~\cite{DBLP:conf/chi/AgrawalABCFHHTK20}. For example, in the collision with the Hot Air Balloon~\cite{Incident-Balloon}, the pilot recklessly overrode warnings that he was entering prohibited airspace. In an accident, we experienced sudden and erratic altitude swings during takeoff, which required the operator to perform an impossibly complex series of actions in order to gain manual control before the sUAS plunged to the ground. \DW must support \texttt{human interaction} testing by allowing users to connect their interactive devices, such as Radio Controllers, to the simulation environment~\cite{hocraffer2017meta,roth2004human}.

\simpleitem{Sensor and Hardware Issues:} While the primary aim of \DW is to test the safe operation of sUAS software applications and deployments, hardware failures, sometimes confounded by environmental factors, are often the primary cause, or a clear contributor, to an accident~\cite{nafeeconfiguring2022}. The most common faults include loss of signal, excessive vibration, compass interference, battery problems~\cite{Incident-Bat1,Incident-HW1}, and motor failures \cite{CONFIG-F1}. Dramatic hardware failures that cause complete and sudden loss of flightworthiness are out of the scope of our current work. However, \DW can test software solutions for detecting, recovering from, and/or preventing hardware faults from occurring. For example, given an event that triggers an RTL failsafe mechanism, can the sUAS return home without colliding with other sUAS? To create this type of test environment, \DW must support \texttt{fault-injection} capabilities that generate sUAS failures at runtime. Test scenarios should define types of failures (e.g., vibration, loss of signal) and their frequencies. While this is out of scope for the current paper, initial work in this area shows that it is feasible to accomplish~\cite{inject-faults1}.

These categories and their associated incidents provide an initial set of guidelines for specifying \DW's configuration requirements, designing a template for test specifications, and identifying a set of monitorable properties.  

\subsection{Configuration Properties}
By analyzing the reported environmentally-related sUAS accidents and incidents, we identified an initial set of relevant contributing factors as depicted in \citetable{factors}. These attributes need to be configurable in order to support meaningful test scenarios. Each parameter is labeled as fully implemented, partially implemented, or planned for future releases. The \emph{Scene} category allows developers to define specific regions of operation that match their actual deployments. Users can also specify various parameters related to sensors and weather conditions. Sensors and hardware faults can be represented by various forms of fault models, and human interactions can be configured by mapping sUAS controls (e.g., the radio controller) to an API associated with each sUAS.

%% file: tables/tab_01_datasources.tex
\begin{table}[!t]
    \vspace{8pt}
    \caption{Overview of sUAS incident sources reported publicly and/or through governmental and regulatory bodies} 
     \vspace{-5pt}
    \addtolength{\tabcolsep}{-3.8pt}
    \centering
 
    \begin{tabular}{ L{3.8cm}L{3.9cm}L{.5cm}}
    
    \toprule
     {\bf Source }& {\bf URL} & {\bf \#}   \\ \midrule
     Aviation Safety Reporting System (ASRS)    & \url{https://asrs.arc.nasa.gov/docs/rpsts/uas.pdf} (ACN: 1599671) &
     50\\ \midrule
     
     Wikipedia collection of incidents &
     \url{https://en.wikipedia.org/wiki/List_of_unmanned_aerial_vehicles-related_incidents} 
     & 90+\\ \midrule
     
     dedrone -- Collection of Worldwide Drone Incidents &
     \url{https://www.dedrone.com/resources/incidents-new/all}&
     70+\\ \midrule
     
     The Center for the Study of the Drone -- Bard College &
     \url{http://dronecenter.bard.edu/drone-incidents}&
     30\\ \midrule
     UK Air Accidents Investigation Branch reports &
     \url{https://www.gov.uk/aaib-reports?keywords=UAS}&
     50\\ \midrule
     
     

    Public Safety Flight -- Report Drone Accident (2019 --2022) & 
     \url{https://reportdroneaccident.com}&50\\

     \bottomrule
    \end{tabular}
    \label{tab:incident_sources}
    \vspace{-13pt}
\end{table}

%% file: tables/tab_02_factors.tex
\begin{table*}[]
    \centering
     \addtolength{\tabcolsep}{-1.0pt}
     \caption{\DW Simulation Requirements with the associated Features and  
corresponding Configurable Properties (\DW Status: \CIRCLE=Implemented; \RIGHTcircle=Partially implemented;  \Circle=Not yet Implemented)}
    \label{tab:factors}
   \begin{tabular}{L{1.65cm}L{4.8cm} L{3.25cm}L{6.5cm}c}
     \toprule
{\bf Category}&{\bf Observed Incidents}&{\bf Feature}& {\bf Configuration Parameters}& \\ \midrule

\multirow{3}{*}{
\begin{minipage}[t]{1.3cm}
    \emph{Scene}\newline(Location \& Terrain)
\end{minipage}
}
 & \multirow{3}{*}{
\begin{minipage}[c]{4.6cm}
sUAS hitting immovable objects, such as buildings or rocks, and interfering with air traffic and humans on the ground~\cite{Incident-Balloon},~\cite{CONFIG-L4},~\cite{CONFIG-L2},~\cite{CONFIG-L3}.
\end{minipage}
}
  &Regions& Region Name, Central coordinates (LLA)& \CIRCLE\\ 
& &Terrain type& Mountainous, Forrested, River, Ocean, Urban.&\RIGHTcircle\\ 

& &Scene& Examples: Air Traffic, Crowd, Ground traffic, accident, building fire, etc.& \Circle\\ \midrule

 \multirow{2}{*}{
 \begin{minipage}[t]{1.6cm}
   \emph{GPS}
\end{minipage}
 }
&  
\multirow{2}{*}{
\begin{minipage}[t]{4.6cm}
sUAS losing GPS signal or experiencing low GPS accuracy during flight~\cite{CONFIG-S1}.
\end{minipage}%
}
& Degree of GPS deprivation & [0-100\%] & \RIGHTcircle\\ 
& & GPS availability plugin & [RandomDisturbance, GNSSModel] & \Circle\\ \midrule

 \multirow{2}{*}{
 \begin{minipage}[c]{1.6cm}
   \emph{sUAS Com\-munication}
\end{minipage}
 }
&  
\multirow{2}{*}{
\begin{minipage}[t]{4.6cm}
sUAS loses the communication link with the remote pilot \cite{CONFIG-S2,CONFIG-S3}.
\end{minipage}%
}
& Fault injection model  & Fault Type (e.g., interference)& \Circle\\ 
& &  Degree of signal deprivation & [0-100\%, duration signal drop]  &  \Circle\\ \midrule

\emph{Weather \& Lighting} &
\multirow{2}{*}{
\begin{minipage}[c]{4.6cm}
sUAS being blown off course due to high winds, operating outside its capabilities, or suffering mechanical issues due to environmental conditions~\cite{CONFIG-W2}.
\end{minipage}
}

& Weather conditions& [Sunny, Cloudy, Raining (heavy, medium, low), Snowing (heavy, medium, low)]& \RIGHTcircle\\ 
&& Lighting conditions& dawn, morning, afternoon, dusk, night& \CIRCLE\\ 
&& Temperature& [Celsius (-30-40), Fahrenheit (-20-120)]& \Circle\\ 
&& Wind& For each range [altitude(upper, lower),wind speed (mph), wind gusts (mph), direction]& \CIRCLE\\ \midrule 

\emph{Human Interactions} &
\multirow{1}{*}{
\begin{minipage}[c]{4.3cm}
sUAS not being operated correctly; issues related to ambiguous UI~\cite{Incident-Balloon}.
\end{minipage}
}
&Mapping RC signals to \DW's exp. inputs & Examples:
Mode switches (e.g., Loiter, Stabilize, RTL, LAND),
Gimbal \& Camera controls& \RIGHTcircle\\ \midrule

\emph{Sensors \& Hardware} &
\multirow{1}{*}{
\begin{minipage}[c]{4.5cm}
sUAS loses power mid-flight; physical damage (e.g.,  motors)~\cite{Incident-Bat1},\cite{Incident-HW1},\cite{CONFIG-F1}.
\end{minipage}
}
&Fault injection model& Fault Model Name,
Fault Types e.g., [Loss-of-signal, Magnetic Interference, Vibration, etc.] & \Circle\\ \bottomrule
    \end{tabular}
\vspace{-15pt}
\end{table*}

%% file: sec_04_architecture.tex
\section{\DW Architecture and Testing Process}
\label{sec:platform}

In this section, we provide a comprehensive overview of the \DW and its components as depicted in \citefig{dwarch}, and the process for executing a test scenario. 
\DW comprises three main components: \emph{(A)} a \emph{Test Scenario Configurator and Generator} that accepts user-defined specifications to configure the environment and test properties and generate variations for fuzzy testing, \emph{(B)} a \emph{Simulation Environment} with sUAS physics engines and a high-fidelity simulated world, and \emph{(C)} a \emph{Runtime Monitoring Environment} that collects and analyzes data during test execution to determine the success or failure of acceptance tests and provide mission analytics visualization for sUAS application diagnosis.

\subsection{Test Scenario Configurator}
Developers configure system-level tests by specifying environmental conditions, sUAS sensor configuration, and test properties based on specified system-level requirements.

\begin{figure}[]
    \centering
    \includegraphics[width=.90\columnwidth]{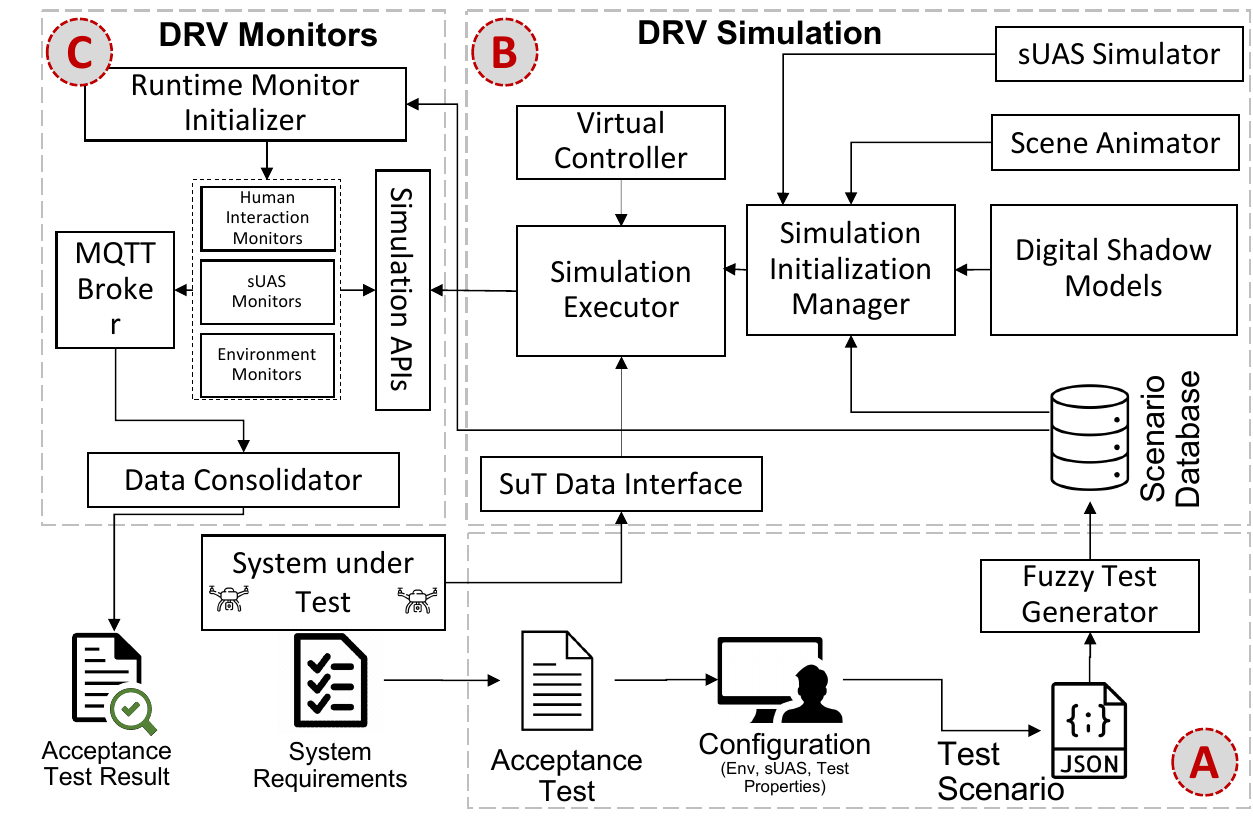}
    \caption{Overview of the high-level \DW~architecture with the 3 main components for test specification, execution, and analysis.}
    \label{fig:dwarch}
    \vspace{-18pt}
\end{figure}

\subsubsection{Configuring the Simulation Environment}
Simulation of realistic environmental conditions is an important part of CPS testing. This includes wind conditions (velocity, direction), geographical regions (urban, densely populated, rural), and time of day (bright sunlight, dark), all of which impact sUAs flight trajectories and mission execution. Additionally, the test scenario scope can be configured according to environmental requirements for a particular mission, such as a drowning victim in a particular river at a specific geographical location or a burning building in a particular area. Based on these inputs, \DW establishes the simulation environment. In the following step, the user configures the characteristics of the sUAS with which they wish to carry out missions.  

\subsubsection{Configuring and Deploying multiple sUAS}
Multi-sUAS systems consist of heterogeneous sUAS, therefore developers must be able to specify the specifications of all sUAS participating in a test scenario.
This includes testing sUAS with specific hardware setups, e.g., sensor configurations and flight missions. As part of this, \DW allows configuring the number of sUAS part of a mission, sensor specifications of each sUAS, and their home geolocations in the simulation environment.  Based on this, \DW deploys multiple sUAS with the defined sensor configurations at the specified location in the simulated environment. Since users are able to ``bring their own sUAS application'', the mission to be expected could be simply planned using an off-the-shelf application, such as QGroundControl~\cite{qground}, or could be developed in a customized sUAS application~\cite{DBLP:conf/chi/AgrawalABCFHHTK20,choi2017enabling,terzi2019swifters}. 




\subsubsection{Specifying Test Properties}
\label{sec:testprops}
Determining mission success is one major aspect in which \DW significantly differs from current sUAS simulation environments and testing processes in which tests are deemed to have passed if the user \emph{observes} correct behavior during simulation. In \DW we replace this ad-hoc process with a structured, and well-defined way of automatically determining if a test (or set of tests) has \texttt{\small PASSED} or \texttt{\small FAILED}.

Success criteria are defined via a set of test properties that must hold true throughout the entire mission in order for the test to be considered to have \texttt{\small PASSED}. The test properties are directly derived from the requirements. Therefore, the third, and last step in the configuration process thus involves configuring the test properties based on safety requirements of the system. For example, specifying the safe distance that two sUAS shall always maintain, or specifying the safe landing spots for each sUAS during the mission.  
\DW leverages its runtime monitoring environment that collects sUAS sensor data at runtime to evaluate whether the safety properties specified by the user hold true during the simulation or not. \DW monitors are further discussed in Section~\ref{sec:runtime_monitors}. 

\subsubsection{Test Scenario Execution \& Fuzzy Test Generation}
\label{sec:fuzzy_implementation}
Before the test scenario is deployed within the simulation engine for execution, \DW provides support for test case fuzzing \cite{sheikhi2022coverage}. This step facilitates testing sUAS applications in uncertain or extreme environmental conditions, without the need to manually create hundreds of test scenarios with slightly different value combinations. Fuzzy testing is intended to determine under which realistic conditions (e.g. max wind velocity) the system-level requirements are satisfied. The test fuzzer component generates multiple copies of user-specified test scenarios and manipulates parameter values to explore sUAS reliability in uncertain environments. Users can specify which environmental configuration to fuzz during test execution. For instance, if a user specifies wind velocity to fuzz, \DW will increase the wind velocity exponentially in each fuzzed scenario to determine unsafe wind velocity for the sUAS to operate in the real world. Additionally, the user must specify the maximum value of the parameter in order to create a termination condition for fuzzy testing. Thus, the \DW architecture and the testing process facilitate the testing of UAS applications under uncertain conditions without requiring the developers to come up with and specify the complex unexpected real-world uncertainties.

\subsection{The Simulated Environment}

\subsubsection{Environment Simulator and Executor}
The environment is structured around accurate 3D geospatial data containing real-world landmarks, such as streets, buildings, bridges,  power lines, and trees. Further, it needs to be augmented by mission-specific objects and phenomena which constitute environments for specific sUAS deployments such as people, vehicles, fires, floods, and avalanches. Therefore, in addition to simulating real-world locations, \DW provides a dedicated \emph{Scene Animator} that  serves as a collection of 3D animations representing typical sUAS mission deployments such as timed simulation of a drowning person in a river or structural fire. 
 The \emph{Simulation Initialization Manager} retrieves user-configured and fuzzed test scenarios from a \emph{Scenario Database} to initialize the simulation environment's initial conditions, including number of sUAS, and weather conditions. 

\subsection{Runtime Monitors and Reports for Analysis}
\label{sec:runtime_monitors}
Runtime monitoring is the process of observing and analyzing the behavior of a software system during its execution~\cite{vierhauser2016reminds}. In order to support critical analysis of simulation results, \DW includes its own runtime monitoring environment to collect data during simulation and check for safety requirement violations. The monitoring component collects data from three sources: sUAS sensors, the environment (e.g., windspeed), and human interactions.

In order to be able to provide ``meaningful'' analysis and Acceptance Test results, it is necessary to provide additional information that goes beyond a simple \texttt{\small PASSED} or \texttt{\small FAILED} for a specific test. In case of undesired behavior, sUAS developers must understand the \emph{how}, \emph{why}, and \emph{where} a particular test scenario failed to satisfy the requirement. For this purpose, \DW takes runtime information and consolidates it for each sUAS in the environment, including any changes to the 3D environment for software analysis purposes. 

%% file: sec_06_evaluation.tex
\section{Empirical Evaluation} 
\label{sec:sim_eval}

To evaluate whether the design of \DW accomplishes the first development objective, DO1, as discussed in Section \ref{sec:overview}, the applicability of \DW to real-world sUAS applications, and its support for testing system-level requirements, we (i) performed a case study following the guidelines described by Runeson and Hoest~\cite{runeson2009guidelines} for two sUAS use cases for which we needed assurances that our DSuT would perform safely in the real world; and (ii) conducted a perception evaluation study~\cite{tory2004human} with sUAS developers asking them to configure a multi-sUAS test scenario based on given requirements. We explore RQ1 defined as follows:

\noindent$\bullet$ \emph{RQ1: }
\emph{How effective is the design of \DW for defining and executing realistic test scenarios for system-level requirements, and how do sUAS developers perceive the overall simulation testing process when using \DW?}

We addressed this RQ in two ways.  First, to evaluate the expressivity and the design of \DW, we applied it to two real-world sUAS application scenarios, creating a series of acceptance tests for each of them, and specifying environmental features and respective monitors relevant to the tests to determine the feasibility of our approach. 

Second, we asked software engineers and sUAS developers to configure their own test scenarios based on given real-world requirements using \DW. After finishing the scenario configuration and analysis of the acceptance test report, we asked the developers a series of follow-up questions to understand how \DW is perceived by end-users and to assess the quality and usefulness of the generated test reports.


To evaluate whether the design of \DW is capable of detecting safety-related concerns in sUAS applications and whether it accomplishes the second objective of \DW development (DO2), we investigate RQ2 as follows:\\
    \noindent $\bullet$ \emph{RQ2:}
    \emph{To what extent can \DW detect and report safety-related issues that occur during tests?}
    
    We addressed this RQ by running a series of simulations in our \DW platform, based on the aforementioned tests, and by seeding faults causing failures in the system to assess if \DW is capable of detecting and documenting them during the simulation.

\subsection{\DW Prototype Implementation}
We implemented an initial prototype to support Pixhawk/PX4 application tests for a number of environmental configuration options and monitorable properties. The configurable properties and features currently supported are marked as implemented, or partially implemented, in~\citetable{factors}.


We developed a web application using the React Web framework that allows users to interact with \DW. The web application has a Wizard that guides users through the process of creating a test scenario to validate a specific requirement, supported by a dashboard that displays the consolidated simulation results for users to analyze. 
Through the web application, users can configure their sUAS, environment and test properties, which are then sent to the backend server written in Python using Flask over HTTP. Using the configuration provided by the user, the back-end server generates the fuzzy test scenario and instantiates the simulation environment. 

Our \DW simulation environment was implemented using the Unreal Engine~\cite{unrealengine}, integrating open-source digital shadow models of the real world by using Cesium for Unreal~\cite{cesium}. We simulated sUAS using AirSim~\cite{shah2018airsim}, an open-source, cross-platform simulator, and used AirSim's APIs to simulate weather conditions in the environment. Furthermore, we implemented runtime monitors as Python modules that collect (using AirSim's APIs) and validate data against the test properties specified by the user. Additionally, interaction data is collected through input devices such as handheld controllers, keyboards, and game controllers. 

\subsection{Drone System Under Test}
Both use cases were enacted using our own semi-autonomous, multi-sUAS system, which includes a Ground Control Station (GCS) supported by a suite of microservices, onboard autonomous processing capabilities, and diverse GUIs to support human interactions. The Onboard Pilot acts as an application layer for the PX4 autopilot stack which includes flight control software and hardware for executing plans. Its internal \textit{State Machine} receives mission specifications and instantiates itself dynamically for the current mission. 


\subsection{Use Case Driven Test Scenarios}
We conducted experiments on two real-world use cases. The first use case represents the deployment of multiple sUAS at an active airbase to collect long-distance imagery of people at diverse pitches and altitudes, and the second use case involved a live search-and-rescue demonstration that we conducted in August 2022 in conjunction with a local Fire Department's water-rescue team. Each use case is described in detail, including its environmental configurations, test properties, and unique requirements.

\noindent $\bullet$ \textbf{UC1 -- Video Collection at Pitch and Range:}
For the first use case, computer vision researchers provided  requirements for collecting aerial images at specific camera pitches and custom flying patterns. The engineering team developed an sUAS application to deploy three sUAS at the test location of an active airbase,  tasked with collecting aerial imagery of people from diverse distances, pitches, and angles, while maintaining minimum separation distance between all sUAS. 

\noindent $\bullet$ \textbf{UC2 -- Search-And-Rescue:}
For our second case, we deployed four Hexacopter sUAS in support of a live search-and-rescue demonstration in collaboration with emergency responders. The public nature of the demonstration, and the fact that it was conducted during the summertime at a crowded beach area, required rigorous testing before deployment. The sUAS were assigned a search area and dispatched to search, utilizing onboard computer vision, and streaming video when a potential sighting was made. In particular, we wanted to simulate the exercise in advance, to evaluate the impact of windy conditions, and to ensure that the critical safety properties held throughout the mission. 


\vspace{-0.2em}
\section{RQ1 -- Test Scenario Definition}
\label{sec:rq1}
\input{sec_06_eval_RQ1}

\section{RQ2 -- Test Scenario Execution}
\label{sec:rq2}
\input{sec_06_eval_RQ2}

%% file: sec_06_eval_RQ1.tex
\vspace{-2pt}
\subsection{Ability to configure real-world requirements}

\input{tables/tab_BRIAR_test}

We applied \DW to the planning and validation of both use cases using our own drone system as the DSuT. We used the \DW interface to configure test scenarios that included digital shadow models of the designated area, no-fly zones, wind, and lighting conditions. Tables \ref{tab:briar-test} and \ref{tab:mc-test} summarize the detailed test configuration for UC1 and UC2. In addition, we identified critical safety requirements for each deployment and translated them into test properties supported by \DW.

\input{tables/tab_MC_test.tex}



The missions flown in \DW were specified entirely using our own DSuT. Tests were then executed successfully and the Acceptance Tests reports were generated for analysis. However, as depicted in Table~\ref{tab:factors}, we have only integrated a subset of the potential configuration parameters in the current prototype. For example, we have not yet integrated any form of sUAS fault injection, GNSS (satellite), or wireless network failures. We, therefore, deliberately did not attempt to configure the environment with these properties and leave their inclusion to the next phase of our work. The configuration of real-world scenarios showed that \DW allows configuring sUAS, environmental settings, and tests based on system-level requirements, simulating sUAS in realistic conditions, and analyzing acceptance test results.

\subsection{Perception-based Evaluation with Practitioners}

We conducted a preliminary evaluation, under an approved study protocol, to assess usability and end-user perception of \DW for creating test scenarios, configuring test properties, and interpreting simulation results.

\textbf{Study Setup \& Execution: } We leveraged our professional networks to recruit five software engineers with experience in requirements engineering and testing. The study was divided into three phases: learning, task performance, and interviewing. In the learning phase, we provided an overview of the objectives of \DW and demonstrated the web interface. In the task performance phase, we assigned a requirement from Table \ref{tab:reqs_user_study} and asked participants to configure test scenarios and analyze the acceptance test report. We used a think-aloud protocol throughout this phase to obtain insights into user opinion and interaction. In the final phase, we asked participants to reflect on the usability of our solution using a questionnaire. Table \ref{tab:study} shows the industrial software engineering experience of study participants and their five-point Likert scale ratings on the ease of test scenario configuration, analyzing acceptance test reports, and examining mission analytics information provided through graphical plots. Each study session took approximately 30 minutes.

\textbf{Results:} The user interface for configuring the multi-sUAS test scenario was intuitive and easy to use for all participants. Participant P3 gave very positive feedback on the platform, suggesting that \DW is a valuable tool for simulation testing. P4, who has been actively working on sUAS applications for three years, emphasized the uniqueness of the \DW capabilities, stating that ``\emph{I have not seen any other sUAS simulation and testing platform that encourages developers to conduct simulation testing based on system-level requirements}''. During the study, participants also identified several issues. P1 reported difficulty specifying the home geolocation of each sUAS, which was time-consuming as they had to copy and paste geocoordinates from Google Maps. P1, P2, and P3 also recommended visualizing the location of multiple sUAS on a 2D map to understand their relative proximity.

During the simulation, participants were interested in observing the flight path of all sUAS in the 3D environment, but found it challenging when the sUAS were far apart. This was due to the view port's inability to cover the entire area occupied by all the sUAS simultaneously, even when adjusted and zoomed in/out. This highlights the need for a solution that supports multi-viewports based on sUAS location, allowing users to observe multiple sUAS even when they are far apart.

After observing the simulation, participants analyzed the generated acceptance test report. P5, with more than 6 years of experience in CPS development, found the auto-generated acceptance test report and accompanying plots depicting deviations in the flight path under varying wind velocities to be a ``\emph{game changer}''. 
P5 found the visualization tools to be extremely helpful in interpreting simulation results and recommended that the feature be integrated into the Continuous Integration (CI) Pipeline of their sUAS development environment. 
P3 recommended that future improvements should include explanations for deviations from flight paths to help developers understand the factors affecting flight paths. 


\input{tables/participants}
\input{tables/table-user-study-reqs}



In response to RQ1, we were able to apply \DW to our two use cases and demonstrate its practical applicability in real-world testing scenarios and integration in a thorough software testing process. We were able to specify the safety requirements, connect our own DSuT with minimal effort, execute test cases, and analyze the results provided by the Acceptance Tests to further improve our application. We also found that developers were able to configure complex test scenarios easily, and acknowledged the usefulness of the Acceptance Test reports. These results demonstrate the practical applicability of our platform in real-world testing and suggest that we achieved our first objective (DO1) of developing \DW as described in Section \ref{sec:overview}. 

\begin{figure}[t]
    \centering
    
    \begin{subfigure}{0.46\columnwidth}
  
    \includegraphics[width=\columnwidth]{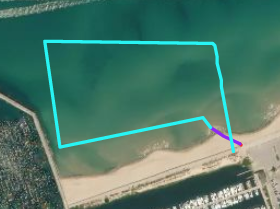}
    \footnotesize{
     Wind speed:~~~~{\it \ul{23 mph}}  \\
     Outcome:~~~~~~~{\it \texttt{PASSED}}
     }
    \end{subfigure}
     \begin{subfigure}{0.40\columnwidth}
    \includegraphics[width=\columnwidth]{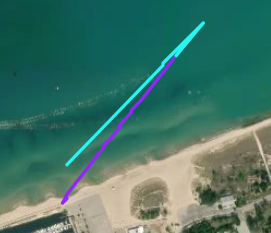}
     \footnotesize{
     Wind speed:~~~~{\it \ul{30 mph}}  \\
     Outcome:~~~~~~~{\it Blown into lake}
     }
    \end{subfigure}
  
    \caption{Flight test passes when winds were $\approx$ 23 mph but unsurprisingly failed at 30 mph (Violating C2.3 described in Table \ref{tab:mc-test})}
    \label{fig:wind}
    \vspace{9pt}

    \begin{subfigure}{\columnwidth}
  \centering
    \includegraphics[width=0.65\columnwidth]{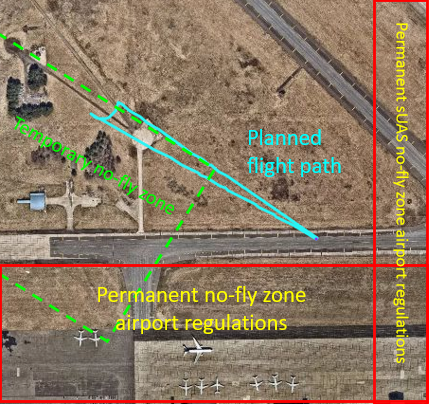}
    \end{subfigure}

       
    
    \caption{Planned Route (blue). However, the flight violated C 1.2 listed in Table \ref{tab:briar-test} by infringing on the temporary no-fly zone (green).  Outcome: \texttt{FAILED}}
    \label{fig:griffiss}
    \vspace{-15pt}
\end{figure}


%% file: tables/tab_BRIAR_test.tex
\begin{table}[t]
    \centering
    \caption{Simulation environment specified for the deployment of sUAS at extreme pitch and range at an active airbase.}
    \label{tab:briar-test}
    \renewcommand{\arraystretch}{0.85}
    \begin{tabular}{|L{1.5cm}|L{6.5cm}|}
    \hline
     \multicolumn{2}{|c|}{\cellcolor{grayhighlight} \bf Example Requirements Under Test}\\ \hline
         \multirow{3}{1.5cm}{Req ID} & \emph{RQ-021}:  When multiple sUAS are in flight, their designated flight regions shall be laterally separated.\\ \hline
          \multirow{2}{1.5cm}{Req ID} &  \emph{RQ-022}: No sUAS shall enter airspace marked as a no-fly zone.\\ \hline
           \multicolumn{2}{c}{\vspace{-0.17cm}} \\ 
        
        \hline
         \multicolumn{2}{|c|}{\cellcolor{grayhighlight}\bf Test Description}\\ \hline
         \multirow{2}{1.5cm}{Test Description}& TST-0023: Six sUAS are deployed across three stations to collect images of actors at various pitches and angles.\\ \hline

           \multicolumn{2}{c}{\vspace{-0.17cm}} \\ \hline
         \multicolumn{2}{|c|}{\cellcolor{grayhighlight}\bf Environment Configuration Parameters} \\ \hline
         \multirow{5}{1.5cm}{Weather}& (\hlcyan{wind=10$\rightarrow$25mph}, $0<ALT<400ft$ AGL)\\
         &
         (\hlcyan{windgusts=0$\rightarrow$10mph} above stable wind,$ALT>400ft$ AGL)\\
         &(precipitation=none, clouds=\hlcyan{light$\rightarrow$heavy})\\ 
         &(timeOfDay=\hlcyan{Dawn$\rightarrow$Dusk})\\\hline
         
         \multirow{2}{1.5cm}{Signal}&(RadioInterference=Light)\\
         &(\hlcyan{Satellites=15-20}, GPSDeadSpots:None)\\ \hline
         \multicolumn{2}{|c|}{\cellcolor{grayhighlight}\bf sUAS Configuration Parameters} \\ \hline
         \multirow{2}{1.5cm}{Location}&(Region=AIRBASE)\\
         & (Central Point=(43.231539, -75.413843))\\ \hline 
         \multicolumn{2}{|c|}{\cellcolor{grayhighlight}\bf Monitors Configuration Parameters} \\ \hline
         \multirow{2}{1.5cm}{Mission Props}& C1.1 Lateral Separation Distance = 10 meters \\
         & C1.2 No Fly Zone Areas, Runway, Taxiway, Actor Zones \\
         
          \hline
         \multicolumn{2}{c}{\vspace{-0.17cm}} \\ \hline
         \multicolumn{2}{|c|}{\cellcolor{grayhighlight}\bf DSuT Flights} \\ \hline
         \multirow{5}{1.5cm}{Test 1}& (BASE-1.flt)          1 Six sUAS deployed to three stations, each station assigned its own image collection task with an associated flight pattern. Simulation lasts 60 minutes to assess changeovers between sUAS. \\\hline
         \end{tabular}
    \vspace{-5pt}
\end{table}

%% file: tables/tab_MC_test.tex
\begin{table}[t]
    \centering
    \caption{Simulation environment specified for the deployment of sUAS for Search-and-Rescue Demonstration at the Beach.}
    \label{tab:mc-test}
    \renewcommand{\arraystretch}{0.95}
    \begin{tabular}{|L{1.5cm}|L{6.5cm}|}
    \hline
         \multicolumn{2}{|c|}{\cellcolor{grayhighlight} \bf Example Requirements Under Test}\\ \hline
         \multirow{2}{1.5cm}{Req ID} & \emph{RQ-034}: sUAS must complete missions successfully at wind speeds of 15mph.\\ \hline
         \multirow{2}{1.5cm}{Req ID} & \emph{RQ-035}: No sUAS shall enter airspace marked as a no-fly zone.\\ \hline
           \multicolumn{2}{c}{\vspace{-0.17cm}} \\ \hline
         \multicolumn{2}{|c|}{\cellcolor{grayhighlight} \bf Test Description}\\ \hline
         \multirow{2}{1.5cm}{Test Description}& TST-0023: sUAS are deployed in a search and rescue activity at the beach\\ \hline
    
         \multicolumn{2}{c}{\vspace{-0.15cm}} \\ \hline
         \multicolumn{2}{|c|}{\cellcolor{grayhighlight} \bf Environment Configuration Parameters} \\ \hline
         \multirow{3}{1.5cm}{Weather}& (wind=15mph, $0<ALT<400ft$ AGL)\\
         &(precipitation=none, clouds=none)\\ 
         &(timeOfDay=Midday)\\\hline
         
         \multirow{2}{1.5cm}{Signal}&(RadioInterference=Light)\\
         &(Satellites=15, GPSDeadSpots:None)\\ \hline
         \multicolumn{2}{|c|}{\cellcolor{grayhighlight} \bf sUAS Configuration Parameters} \\ \hline
         \multirow{2}{1.5cm}{Location}&(Region=BEACH)\\
         & (Central Point=(42.207762, -86.393095))\\ 
         \hline 
         \multicolumn{2}{|c|}{\cellcolor{grayhighlight} \bf Monitors Configurations Parameters} \\ \hline
         \multirow{3}{1.5cm}{Mission Props}& C2.1 No Fly Zone = Beach Area\\ 
             & C2.2 Safe Landing Spots = Anywhere on Ground except water.\\
          & C2.3 Drift $<$ 10\% when Wind $>=$ 23mph.\\
        
          \hline
         \multicolumn{2}{c}{\vspace{-0.15cm}} \\ \hline
         \multicolumn{2}{|c|}{\cellcolor{grayhighlight} \bf DSuT Flights} \\ \hline
         \multirow{3}{1.5cm}{Test 1}&(BEACH-1.flt)~Four sUAS deployed for search and rescue. They fly over the water and the dunes to search for the lost child. Simulation lasts 20 minutes. \\\hline
         \end{tabular}
        \vspace{-15pt}
\end{table}

%% file: tables/participants.tex



\begin{table}[t!]
\centering
\caption{Participant Details and their responses on 5 point likert scale}
\begin{tabular}{p{0.15cm}R{0.80cm}L{0.5cm} | L{1.55cm}L{1.30cm}L{1.90cm}}
\toprule
\multirow{2}{*}{\textbf{ID}} & \multirow{1}{*}{\textbf{Exp.}}&\multirow{1}{*}{\textbf{Req}} & \multicolumn{3}{c}{\textbf{Ratings}} \\
& \textbf{( Yrs.)} &\textbf{Id}& \textbf{Scenario Config} & \textbf{Report Analysis} & \textbf{Flight Path Vizualization} \\
\midrule
P1 & 2 & 301 &3 (Moderate) & 5 (V. Easy) & 4 (V. Helpful) \\
P2 & 7 &303& 4 (Easy) & 5 (V. Easy) & 5 (Ext. Helpful) \\
P3 & 8 &303& 4 (Easy) & 4 (Easy) & 4 (V. Helpful) \\
P4 & 7 &301& 5 (V. Easy) & 4 (Easy) & 4 (V. Helpful) \\
P5 & 13 &302& 4 (Easy) &  4 (Easy) & 5 (Ext. Helpful) \\
\bottomrule

\end{tabular}

\label{tab:study}
\vspace{-5pt}
\end{table}

%% file: tables/table-user-study-reqs.tex

\begin{table}[]
\centering
\caption{Description of Requirements Assigned to Participants for Testing}
\label{tab:reqs_user_study}
\begin{tabular}{cp{7.0cm}}
\toprule
\textbf{Req ID} & \textbf{Description} \\
\midrule
301 & Two sUAS fly a circular and square flight mission in windy conditions, avoiding collisions and drifting more than 10m. \\ \midrule
302 & Two sUAS complete missions in windy conditions, maintaining 5m separation and drifting less than 5m. \\ \midrule
303 & Two sUAS complete missions in windy conditions, drifting less than 15m. \\ \bottomrule
\end{tabular}

     \vspace{-15pt}
\end{table}

%% file: sec_06_eval_RQ2.tex

To address RQ2 we executed a series of tests for each of the use cases. We deliberately included test cases that we expected to fail, in order to evaluate \DW's ability to raise errors appropriately. Here, we report one example from each of the use cases. In \citefig{wind} we plot flight logs from simulations in winds of 23mph and 30mph. The first test passed, whilst the second one (at 30mph) failed when the sUAS blew away into the lake. Second, in Figure~\ref{fig:griffiss} we show several no-fly zones at the airport. In this example, a flight triggered a \texttt{\small FAILED} test because the flight path flew over a temporary no-fly zone violating C1.2. These and many other examples demonstrated that \DW was able to accurately differentiate between \texttt{\small PASSED} and \texttt{\small FAILED} test cases.

\begin{figure*}[]
  \begin{minipage}[htbp]{.25\linewidth}
    \includegraphics[width=\linewidth]{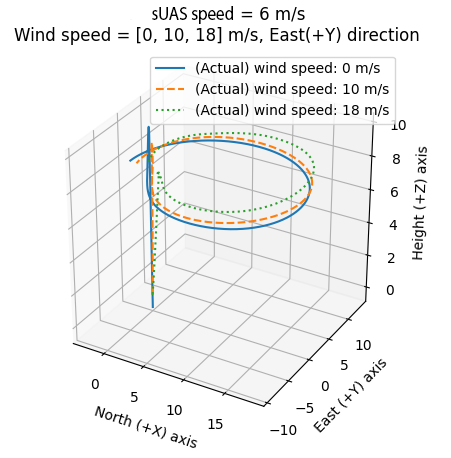}
    \label{fig:drone_speed_6}
  \end{minipage}\hfill
  \begin{minipage}[htbp]{.25\linewidth}
    \includegraphics[width=\linewidth]{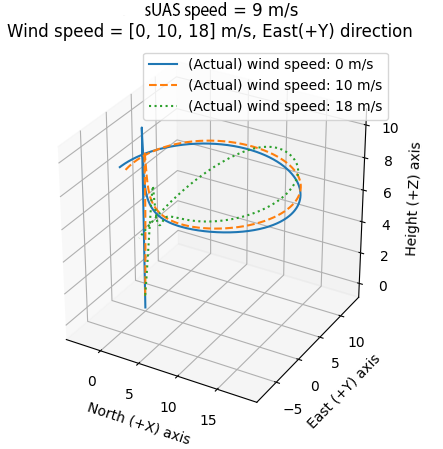}
    \label{fig:drone_speed_9}
  \end{minipage}\hfill
  \begin{minipage}[htbp]{.25\linewidth}
   \includegraphics[width=\linewidth]{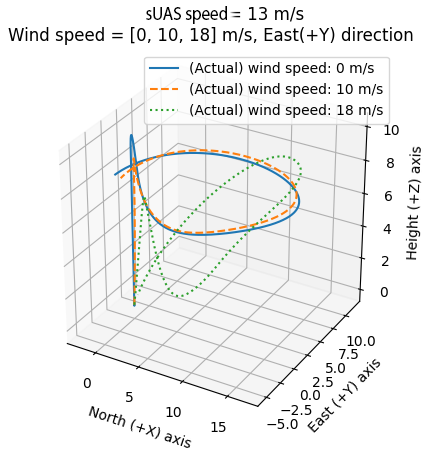}
    \label{fig:drone_speed_13}
  \end{minipage}%
  \vspace{-12pt}
  \caption{Part of generated Acceptance Test report showing the operating boundary of sUAS application across varying sUAS velocity and wind velocity}
  \label{fig:fuzz_results}
  \vspace{-18pt}
\end{figure*}

\vspace{-2pt}

\textbf{Identification of Operating Boundaries:} Incorporating fuzzy testing in the framework enables \DW to automatically compare simulation results across test scenarios, thereby testing an entire range of environmental conditions for a single requirement. To evaluate the practical application of this, we used Airsim's default algorithm to fly a UAV in a circular trajectory at different velocities, created test scenarios, and configured \DW to fuzz the wind velocity, with a maximum wind velocity of 18 meters/s. The sUAS flew at velocities of 6, 9, and 13 meters/s under no wind, with additional simulations conducted at 10 and 18 m/s wind velocity. Results were analyzed to determine the sUAS's ability to withstand varying wind conditions. Figure~\ref{fig:fuzz_results} shows a series of plots generated by \DW monitoring environment to visualize how the increasing wind velocity impacts the sUAS's flight path. These auto-generated plots provided insight into the ability of sUAS to fly circular missions under varying wind conditions. 

In response to RQ2, we found that the sUAS monitoring environment is capable of reporting violations when the sUAS breaches any of the user-configured test properties, as demonstrated in our two use case executions. Additionally, our findings reveal that fuzzing the environmental configuration is effective in identifying operational constraints of the sUAS application. Comparing simulation results from the fuzzed test scenario to the actual test scenario provided valuable insights into the safe behavior of sUAS under varying environmental conditions. These findings also suggest that we achieved our second objective (DO2) of developing \DW.

%% file: sec_07_threats.tex
\section{Threats to Validity}
\label{sec:threats}
In this section, we discuss threats to validity of our study and execution of the test scenarios.


\noindent$\bullet$ \textbf{Construct Validity} refers to how accurately a test measures the concept it was designed to evaluate. In the case of the perception study, participants came from our own network, and their evaluation of \DW's usability could have been biased to provide positive feedback; however, their responses were supported by qualitative responses which provided clear rationales supporting their positions.

\noindent$\bullet$ \textbf{Internal Validity} describes threats that could potentially cause the observed effects besides the independent variable.  In the case of RQ2, we evaluated the extent to which \DW could detect safety-related issues. However, here we assume that the underlying physics engine provides a high-fidelity simulation and accurate proxy of the actual physical sUAS. If the physics of the engine and that of the drone differ considerably, then certain types of faults that occur in the real world will not be detected in simulation. Future versions of \DW will address this by supporting fidelity tests and providing an interface to configure the actual physics engine.

\noindent$\bullet$ \textbf{External Validity} describes threats to the generalizability of results. 
First, while the two use cases used for the evaluation represent complex and real-world applications, exhibiting safety-critical aspects, our evaluation is based on applying \DW to only one DSuT and one type of flight controller. Therefore, additional case studies are required to evaluate generalizability and broader applicability of \DW. As part of this effort, we specifically provide APIs to facilitate tests with diverse sUAS and additional types of missions. Second, while the feedback we received from the five highly experienced software engineers is valuable, they may not necessarily represent the opinions and preferences of a broader range of the development community.

%% file: sec_08_related_work.tex
\section{Related Work}
\label{sec:relwork}





\noindent\emph{\textbf{sUAS Simulation}:} 
Simulation is  widely used in AV research, with many open-source simulators for self-driving cars, such as TORCS \cite{wymann2000torcs} and CARLA \cite{dosovitskiy2017carla}. However, in the sUAS domain, rich simulation environments are far more limited, particularly, with regard to the application-specific scenarios that can be defined and properties that can be monitored during a simulation run. Dronology \cite{cleland2018dronology} is a centralized, multi-sUAS platform with built-in run-time monitors. It has been replaced by DroneResponse which uses a distributed model that supports greater autonomy \cite{nafeeconfiguring2022}. Neither of these platforms provides a structured testing harness to specify and validate the specific sUAS behavior, nor do they consider the complexities of the real world during simulation. With regards to ``pure'' simulation tools, Gazebo~\cite{gazebo}, a 3D robotic simulation platform, and AirSim~\cite{shah2018airsim} facilitate basic sUAS testing. However, these tools alone are limited in terms of providing a structured and well-defined testing environment. Afzal~\etal~\cite{afzal2021gzscenic}, released Gzscenic that lets developers specify various environment elements using domain language and automatically arranges them in Gazebo for simulation. However, there are limitations to the domain language in terms of specifying more complex and 3D realistic landscapes. 
In contrast, our current approach facilitates the testing of multi-sUAS applications in complex scenarios by simulating realistic 3D environmental conditions, automating fuzzy testing, and generating acceptance test results to facilitate debugging.

\noindent\emph{\textbf{sUAS V\&V Activities}:}  
To validate the impact of adverse weather conditions on flight dynamics, researchers have built climate-controlled facilities \cite{scanavino2019new}. However, testing in climate-controlled facilities is expensive and difficult. Grigoropoulos and Lalis~\cite{grigoropoulos2020simulation} describe a simulation environment that allows testing and execution of sUAS applications using digital twin representation of sUAS to detect changes and deviations from requirements. Similar to our work on sUAS testing, Schmittle~\etal~\cite{schmittle2018openuav} have proposed OpenUAV, a testbed for UAV testing, focusing on UAV education and research activities.  OpenUAV provides a dedicated frontend interface and containerized architecture to facilitate UAV testing. Similarly, StellaUAV \cite{schmidt2022stellauav} lets developers build test scenarios using simplistic terrain types, obstacle types, and weather conditions. Khatiri \etal~\cite{khatiri2023simulation} created SURREALIST, a tool that uses real UAV flight data to generate test cases in a simulated environment. However, while all these approaches do provide testing capabilities for sUAS applications, with \DW we focus specifically on enabling mission-specific tests with a particular focus on sUAS interaction with the digital shadows of real-world in realistic scenarios.


AI capabilities have enabled sUAS systems to make autonomous decisions. However, human interventions in sUAS autonomy remain necessary to ensure system safety \cite{agrawal2020model}. Recently, Cleland-Huang~\etal proposed the MAPE-K$_{HMT}$ framework \cite{cleland2022extending}, which augments the traditional MAPE-K loop with Human-Machine Teaming (HMT) requirements \cite{vierhauser2021hazard}. Current sUAS simulation tools lack emphasis on human-sUAS interaction testing. In contrast, \DW proposed architecture monitors human inputs and facilitates human-sUAS control switching testing during simulation. 

%% file: main.bbl
\begin{thebibliography}{10}
\providecommand{\url}[1]{#1}
\csname url@samestyle\endcsname
\providecommand{\newblock}{\relax}
\providecommand{\bibinfo}[2]{#2}
\providecommand{\BIBentrySTDinterwordspacing}{\spaceskip=0pt\relax}
\providecommand{\BIBentryALTinterwordstretchfactor}{4}
\providecommand{\BIBentryALTinterwordspacing}{\spaceskip=\fontdimen2\font plus
\BIBentryALTinterwordstretchfactor\fontdimen3\font minus
  \fontdimen4\font\relax}
\providecommand{\BIBforeignlanguage}[2]{{%
\expandafter\ifx\csname l@#1\endcsname\relax
\typeout{** WARNING: IEEEtran.bst: No hyphenation pattern has been}%
\typeout{** loaded for the language `#1'. Using the pattern for}%
\typeout{** the default language instead.}%
\else
\language=\csname l@#1\endcsname
\fi
#2}}
\providecommand{\BIBdecl}{\relax}
\BIBdecl

\bibitem{ilachinski2017artificial}
A.~Ilachinski, ``Artificial intelligence and autonomy: Opportunities and
  challenges,'' 2017.

\bibitem{castellanos2019modular}
J.~H. Castellanos and J.~Zhou, ``A modular hybrid learning approach for
  black-box security testing of {CPS},'' in \emph{Proc. of the Int'l Conference
  on Applied Cryptography and Network Security}.\hskip 1em plus 0.5em minus
  0.4em\relax Springer, 2019, pp. 196--216.

\bibitem{zhang2016understanding}
M.~Zhang, B.~Selic, S.~Ali, T.~Yue, O.~Okariz, and R.~Norgren, ``Understanding
  uncertainty in cyber-physical systems: a conceptual model,'' in \emph{Proc.
  of the European Conf. on Modelling Foundations and Applications}.\hskip 1em
  plus 0.5em minus 0.4em\relax Springer, 2016, pp. 247--264.

\bibitem{ali2015u}
S.~Ali and T.~Yue, ``{U-Test}: Evolving, modelling and testing realistic
  uncertain behaviours of cyber-physical systems,'' in \emph{Proc. of the 8th
  Int'l Conference on Software Testing, Verification and Validation
  (ICST)}.\hskip 1em plus 0.5em minus 0.4em\relax IEEE, 2015, pp. 1--2.

\bibitem{antkiewicz2020modes}
M.~Antkiewicz, M.~Kahn, M.~Ala, K.~Czarnecki, P.~Wells, A.~Acharya, and
  S.~Beiker, ``Modes of automated driving system scenario testing: Experience
  report and recommendations,'' \emph{SAE Technical Papers}, vol. 2020, no.
  April, 2020.

\bibitem{nasa-aviation}
J.~Cleland-Huang, N.~Chawla, M.~Cohen, M.~N.~A. Islam, U.~Sinha, L.~Spirkovska,
  Y.~Ma, S.~Purandare, and M.~T. Chowdhury, ``Towards real-time safety analysis
  of small unmanned aerial systems in the national airspace,'' in \emph{AIAA
  AVIATION 2022 Forum}, 2022.

\bibitem{CONFIG-W1}
{Gov.UK -- Air Accidents Investigation Branch}, ``Loss of control due to high
  winds and software error,''
  \url{https://www.gov.uk/aaib-reports/aaib-investigation-to-aeryon-skyranger-r60-uas-sr9112798},
  [Accessed on 2023-01-01].

\bibitem{CONFIG-W2}
{Gov.UK - Air Accidents Investigation Branch}, ``Loss of power to aircraft in
  turbulent conditions,''
  \url{https://www.gov.uk/aaib-reports/aaib-investigation-to-yuneec-h520-uas-registration-n-a-240420},
  [Accessed on 2023-01-01].

\bibitem{anda2019arithmetic}
A.~A. Anda and D.~Amyot, ``Arithmetic semantics of feature and goal models for
  adaptive cyber-physical systems,'' in \emph{Proc. of the 27th Int'l
  Requirements Engineering Conference}.\hskip 1em plus 0.5em minus 0.4em\relax
  IEEE, 2019, pp. 245--256.

\bibitem{jung2009real}
D.~Jung, J.~Ratti, and P.~Tsiotras, ``Real-time implementation and validation
  of a new hierarchical path planning scheme of {UAVs} via hardware-in-the-loop
  simulation,'' \emph{Journal of Intelligent and Robotic Systems}, vol.~54,
  no.~1, pp. 163--181, 2009.

\bibitem{zheng2015perceptions}
X.~Zheng, C.~Julien, M.~Kim, and S.~Khurshid, ``{Perceptions on the State of
  the Art in Verification and Validation in Cyber-Physical Systems},''
  \emph{IEEE Systems Journal}, vol.~11, no.~4, pp. 2614--2627, 2015.

\bibitem{dronology}
{Dronology}, ``{Research Incubator and Dataset},''
  \url{https://dronology.info}, 2020, [Accessed on 2023-01-01].

\bibitem{fernando2013modelling}
H.~Fernando, A.~De~Silva, M.~De~Zoysa, K.~Dilshan, and S.~Munasinghe,
  ``{Modelling, simulation and implementation of a quadrotor UAV},'' in
  \emph{Proc. of the 8th Int'l Conference on Industrial and Information
  Systems}.\hskip 1em plus 0.5em minus 0.4em\relax IEEE, 2013, pp. 207--212.

\bibitem{rodriguez2015design}
V.~Rodriguez-Fernandez, H.~D. Men{\'e}ndez, and D.~Camacho, ``{Design and
  development of a lightweight multi-UAV simulator},'' in \emph{Proc. of the
  2nd Int'l Conference on Cybernetics}.\hskip 1em plus 0.5em minus 0.4em\relax
  IEEE, 2015, pp. 255--260.

\bibitem{gazebo}
{Open Robotics}, ``{Gazebo } {Simulator},'' \url{https://gazebosim.org}, 2021,
  [Accessed on 2023-01-01].

\bibitem{shah2018airsim}
S.~Shah, D.~Dey, C.~Lovett, and A.~Kapoor, ``Airsim: High-fidelity visual and
  physical simulation for autonomous vehicles,'' in \emph{Field and Service
  Robotics}.\hskip 1em plus 0.5em minus 0.4em\relax Springer, 2018, pp.
  621--635.

\bibitem{koenig2004design}
N.~Koenig and A.~Howard, ``Design and use paradigms for gazebo, an open-source
  multi-robot simulator,'' in \emph{Proc. of the 2004 Int'l Conference on
  Intelligent Robots and Systems}, vol.~3.\hskip 1em plus 0.5em minus
  0.4em\relax IEEE, 2004, pp. 2149--2154.

\bibitem{abbyasov2020automatic}
B.~Abbyasov, R.~Lavrenov, A.~Zakiev, K.~Yakovlev, M.~Svinin, and E.~Magid,
  ``Automatic tool for gazebo world construction: from a grayscale image to a
  3d solid model,'' in \emph{Proc. of the 2020 IEEE Int'l Conference on
  Robotics and Automation}.\hskip 1em plus 0.5em minus 0.4em\relax IEEE, 2020,
  pp. 7226--7232.

\bibitem{madaan_airsim_2020}
R.~Madaan, N.~Gyde, S.~Vemprala, M.~Brown, K.~Nagami, T.~Taubner,
  E.~Cristofalo, D.~Scaramuzza, M.~Schwager, and A.~Kapoor, ``{AirSim} drone
  racing lab,'' in \emph{Proc. of the {NeurIPS} 2019 Competition and
  Demonstration Track}.\hskip 1em plus 0.5em minus 0.4em\relax {PMLR}, 2020,
  pp. 177--191, {ISSN}: 2640-3498.

\bibitem{wieringa2014design}
R.~J. Wieringa, \emph{Design science methodology for information systems and
  software engineering}.\hskip 1em plus 0.5em minus 0.4em\relax Springer, 2014.

\bibitem{nafeeconfiguring2022}
M.~N. Al~Islam, M.~T. Chowdhury, A.~Agrawal, M.~Murphy, R.~Mehta,
  D.~Kudriavtseva, and J.~Cleland-Huang, ``Configuring mission-specific
  behavior in a product line of collaborating small unmanned aerial systems,''
  \emph{Journal of Systems and Software}, 2022.

\bibitem{ayerdi2020towards}
J.~Ayerdi, A.~Garciandia, A.~Arrieta, W.~Afzal, E.~Enoiu, A.~Agirre,
  G.~Sagardui, M.~Arratibel, and O.~Sellin, ``Towards a taxonomy for eliciting
  design-operation continuum requirements of cyber-physical systems,'' in
  \emph{Proc. of the 28th Int'l Requirements Engineering Conference}.\hskip 1em
  plus 0.5em minus 0.4em\relax IEEE, 2020, pp. 280--290.

\bibitem{abbaspour2015survey}
S.~Abbaspour~Asadollah, R.~Inam, and H.~Hansson, ``A survey on testing for
  {Cyber Physical System},'' in \emph{Proc. of the IFIP Int'l Conference on
  Testing Software and Systems}.\hskip 1em plus 0.5em minus 0.4em\relax
  Springer, 2015, pp. 194--207.

\bibitem{zhou2018review}
X.~Zhou, X.~Gou, T.~Huang, and S.~Yang, ``Review on testing of cyber physical
  systems: Methods and testbeds,'' \emph{IEEE Access}, vol.~6, pp.
  52\,179--52\,194, 2018.

\bibitem{vierhauser2016reminds}
M.~Vierhauser, R.~Rabiser, P.~Gr{\"u}nbacher, K.~Seyerlehner, S.~Wallner, and
  H.~Zeisel, ``Reminds: A flexible runtime monitoring framework for systems of
  systems,'' \emph{Journal of Systems and Software}, vol. 112, pp. 123--136,
  2016.

\bibitem{meier2015px4}
L.~Meier, D.~Honegger, and M.~Pollefeys, ``Px4: A node-based multithreaded open
  source robotics framework for deeply embedded platforms,'' in \emph{Proc. of
  the 2015 IEEE Int'l Conference on Robotics and Automation}.\hskip 1em plus
  0.5em minus 0.4em\relax IEEE, 2015, pp. 6235--6240.

\bibitem{px4}
PX4, ``{PX4 Autopilot},'' \url{https://docs.px4.io/main/en}, 2022, [Accessed on
  2023-01-01].

\bibitem{ardupilot}
{Ardupilot}, ``{Ardupilot Open-Source Flight Controller},''
  \url{https://ardupilot.org}, 2022, [Accessed on 2023-01-01].

\bibitem{nguyen2022bedivfuzz}
H.~L. Nguyen and L.~Grunske, ``{BeDivFuzz}: integrating behavioral diversity
  into generator-based fuzzing,'' in \emph{Proc. of the 44th Int'l Conference
  on Software Engineering}, 2022, pp. 249--261.

\bibitem{sun2022lawbreaker}
Y.~Sun, C.~M. Poskitt, J.~Sun, Y.~Chen, and Z.~Yang, ``{LawBreaker}: An
  approach for specifying traffic laws and fuzzing autonomous vehicles,'' in
  \emph{Proc. of the 37th IEEE/ACM Int'l Conference on Automated Software
  Engineering}, 2022, pp. 1--12.

\bibitem{Popovic2020}
M.~Popovi{\'{c}}, T.~Vidal-Calleja, G.~Hitz, J.~J. Chung, I.~Sa, R.~Siegwart,
  and J.~Nieto, ``{An informative path planning framework for UAV-based terrain
  monitoring},'' \emph{Autonomous Robots}, vol.~44, no.~6, pp. 889--911, Jul
  2020.

\bibitem{kang2018damage}
D.~Kang and Y.-J. Cha, ``{Damage detection with an autonomous UAV using deep
  learning},'' in \emph{Sensors and Smart Structures Technologies for Civil,
  Mechanical, and Aerospace Systems 2018}, vol. 10598.\hskip 1em plus 0.5em
  minus 0.4em\relax SPIE, 2018, pp. 7--14.

\bibitem{DBLP:journals/jcm/DuangsuwanM21}
S.~Duangsuwan and M.~M. Maw, ``Comparison of path loss prediction models for
  {UAV} and iot air-to-ground communication system in rural precision farming
  environment,'' \emph{Journal of Communications}, vol.~16, no.~2, pp. 60--66,
  2021.

\bibitem{DBLP:conf/icse-wain/AbrahamCBVAISC21}
S.~J. Abraham, Z.~Carmichael, S.~Banerjee, R.~G. VidalMata, A.~Agrawal,
  M.~N.~A. Islam, W.~J. Scheirer, and J.~Cleland{-}Huang, ``Adaptive autonomy
  in human-on-the-loop vision-based robotics systems,'' in \emph{Proc. of the
  1st {IEEE/ACM} Workshop on {AI} Engineering - Software Engineering for
  AI}.\hskip 1em plus 0.5em minus 0.4em\relax {IEEE}, 2021, pp. 113--120.

\bibitem{vierhauser2021hazard}
M.~Vierhauser, M.~N.~A. Islam, A.~Agrawal, J.~Cleland-Huang, and J.~Mason,
  ``Hazard analysis for human-on-the-loop interactions in suas systems,'' in
  \emph{Proc. of the 29th ACM Joint Meeting on European Software Engineering
  Conf. and Symposium on the Foundations of Software Engineering}, 2021, pp.
  8--19.

\bibitem{DBLP:journals/corr/abs-2207-08857}
M.~N.~A. Islam, Y.~Ma, P.~A. Granadeno, N.~V. Chawla, and J.~Cleland{-}Huang,
  ``{RESAM:} requirements elicitation and specification for deep-learning
  anomaly models with applications to {UAV} flight controllers,'' \emph{CoRR},
  vol. abs/2207.08857, 2022.

\bibitem{Incident-Balloon}
{Teton Valley News}, ``First-ever recorded drone-hot air balloon collision
  prompts safety conversation,''
  \url{https://apnews.com/article/44695c678bdb44f19172cae4ef842381}, 2018,
  [Accessed on 2023-01-01].

\bibitem{CONFIG-L4}
{dronexl.co}, ``Goose hunting with drones results in crash with dji mavic,''
  \url{https://dronexl.co/2021/02/05/goose-hunting-with-drones}, [Accessed on
  2023-01-01].

\bibitem{CONFIG-L2}
{BBC News}, ``Drone hit newly erected crane during kent site survey,''
  \url{https://www.bbc.com/news/technology-42718824}, [Accessed on 2023-01-01].

\bibitem{CONFIG-L3}
Longmontleader, ``Investigators: Table mountain fire caused by drone crash,''
  \url{https://www.longmontleader.com/local-sports/investigators-table-mountain-fire-caused-by-drone-crash-5287986},
  [Accessed on 2023-01-01].

\bibitem{CONFIG-S1}
Forbes, ``Drone crash due to gps interference in u.k. raises safety
  questions,''
  \url{https://www.forbes.com/sites/davidhambling/2020/08/10/investigation-finds-gps-interference-caused-uk-survey-drone-crash/?sh=1400067f534a},
  [Accessed on 01/08/2022].

\bibitem{CONFIG-S2}
{USA}, ``{AirForceTimes},''
  \url{https://www.airforcetimes.com/news/your-air-force/2019/02/13/that-second-predator-crash-in-four-days-back-in-2017-was-due-to-signal-loss},
  [Accessed on 2023-01-01].

\bibitem{CONFIG-S3}
{Public Intelligence}, ``Lost-links and mid-air collisions: The problems with
  domestic drones,''
  \url{https://publicintelligence.net/the-problems-with-domestic-drones}, 2012,
  [Accessed on 2023-01-01].

\bibitem{Incident-Bat1}
{Gov.UK - Air Accidents Investigation Branch}, ``{AAIB investigation to DJI
  Matrice 210 RTK},''
  \url{https://www.gov.uk/aaib-reports/aaib-investigation-to-dji-matrice-210-rtk-uas-registration-n-a},
  2021, [Accessed on 2023-01-01].

\bibitem{Incident-HW1}
reportdroneaccident.com, ``{Drone Incident Report},''
  \url{https://reportdroneaccident.com/2242/tarot-iron-man-1000-accident-2021-10-19},
  2022, [Accessed on 2023-01-01].

\bibitem{CONFIG-F1}
{Gov.UK -- Air Accidents Investigation Branch}, ``Loss of power due to motor
  failure,''
  \url{https://www.gov.uk/aaib-reports/aaib-investigation-to-dji-matrice-210-uas-registration-n-a-3-march-2019},
  [Accessed on 2023-01-01].

\bibitem{qground}
{QGroundControl – Drone Control}, ``{QGroundControl},''
  \url{http://qgroundcontrol.com}, 2022, [Accessed on 2023-01-01].

\bibitem{mplanner}
{Ardupilot}, ``{MissionPlanner},'' \url{https://ardupilot.org/planner}, 2022,
  [Accessed on 2023-01-01].

\bibitem{Incident-RTLOverride}
{CBS Bay Area}, ``{Pilot Of Drone That Nearly Hit CHP Helicopter Says It Was On
  Autopilot},''
  \url{https://sanfrancisco.cbslocal.com/2015/12/17/drone-near-miss-chp-helicopter-martinez-owen-ouyang-apology-autopilot},
  2015, [Accessed on 2023-01-01].

\bibitem{gao2021weather}
M.~Gao, C.~H. Hugenholtz, T.~A. Fox, M.~Kucharczyk, T.~E. Barchyn, and P.~R.
  Nesbit, ``Weather constraints on global drone flyability,'' \emph{Scientific
  reports}, vol.~11, no.~1, pp. 1--13, 2021.

\bibitem{lundby2019towards}
T.~Lundby, M.~P. Christiansen, and K.~Jensen, ``Towards a weather analysis
  software framework to improve uas operational safety,'' in \emph{Proc. of the
  2019 Int'l Conference on Unmanned Aircraft Systems}.\hskip 1em plus 0.5em
  minus 0.4em\relax IEEE, 2019, pp. 1372--1380.

\bibitem{hamzeh2022improving}
Y.~Hamzeh, A.~Mohammadi, and S.~A. Rawashdeh, ``Improving the performance of
  automotive vision-based applications under rainy conditions,'' \emph{IET
  Image Processing}, 2022.

\bibitem{DBLP:conf/chi/AgrawalABCFHHTK20}
A.~Agrawal, S.~J. Abraham, B.~Burger, C.~Christine, L.~Fraser, J.~M. Hoeksema,
  S.~Hwang, E.~Travnik, S.~Kumar, W.~J. Scheirer, J.~Cleland{-}Huang,
  M.~Vierhauser, R.~Bauer, and S.~Cox, ``The next generation of human-drone
  partnerships: Co-designing an emergency response system,'' in \emph{Proc. of
  the 2020 {CHI} Conference on Human Factors in Computing Systems}, 2020, pp.
  1--13.

\bibitem{hocraffer2017meta}
A.~Hocraffer and C.~S. Nam, ``A meta-analysis of human-system interfaces in
  unmanned aerial vehicle (uav) swarm management,'' \emph{Applied Ergonomics},
  vol.~58, pp. 66--80, 2017.

\bibitem{roth2004human}
E.~M. Roth, M.~L. Hanson, C.~Hopkins, V.~Mancuso, and G.~L. Zacharias, ``Human
  in the loop evaluation of a mixed-initiative system for planning and control
  of multiple uav teams,'' in \emph{Proc. of the Human Factors and Ergonomics
  Society Annual Meeting}, vol.~48, no.~3.\hskip 1em plus 0.5em minus
  0.4em\relax SAGE Publications Sage CA: Los Angeles, CA, 2004, pp. 280--284.

\bibitem{inject-faults1}
G.~Neimiec, ``Design and implementation of a fault injection prototype on an
  autonomous vehicle simulator,'' Bachelor Thesis, Universidade Federal Do Rio
  Grande Do Sul, 2021.

\bibitem{choi2017enabling}
S.-C. Choi, N.-M. Sung, J.-H. Park, I.-Y. Ahn, and J.~Kim, ``{Enabling drone as
  a service: OneM2M-based UAV/drone management system},'' in \emph{Proc. of the
  Ninth Int'l Conference on Ubiquitous and Future Networks}.\hskip 1em plus
  0.5em minus 0.4em\relax IEEE, 2017, pp. 18--20.

\bibitem{terzi2019swifters}
M.~Terzi, A.~Anastasiou, P.~Kolios, C.~Panayiotou, and T.~Theocharides,
  ``Swifters: A {multi-UAV} platform for disaster management,'' in \emph{Proc.
  of the 2019 Int'l Conference on Information and Communication Technologies
  for Disaster Management}.\hskip 1em plus 0.5em minus 0.4em\relax IEEE, 2019,
  pp. 1--7.

\bibitem{sheikhi2022coverage}
S.~Sheikhi, E.~Kim, P.~S. Duggirala, and S.~Bak, ``Coverage-guided fuzz testing
  for cyber-physical systems,'' in \emph{Proc. of the 13th ACM/IEEE Int'l
  Conference on Cyber-Physical Systems}.\hskip 1em plus 0.5em minus 0.4em\relax
  IEEE, 2022, pp. 24--33.

\bibitem{runeson2009guidelines}
P.~Runeson and M.~H{\"o}st, ``Guidelines for conducting and reporting case
  study research in software engineering,'' \emph{Empirical software
  engineering}, vol.~14, no.~2, pp. 131--164, 2009.

\bibitem{tory2004human}
M.~Tory and T.~Moller, ``Human factors in visualization research,'' \emph{IEEE
  Transactions on Visualization and Computer Graphics}, vol.~10, no.~1, pp.
  72--84, 2004.

\bibitem{unrealengine}
\BIBentryALTinterwordspacing
{Epic Games}, ``Unreal engine,'' [Accessed on 2023-01-01]. [Online]. Available:
  \url{https://www.unrealengine.com}
\BIBentrySTDinterwordspacing

\bibitem{cesium}
Cesium, ``Cesium for unreal,''
  \url{https://cesium.com/platform/cesium-for-unreal}, [Accessed on
  2023-01-01].

\bibitem{wymann2000torcs}
B.~Wymann, E.~Espi{\'e}, C.~Guionneau, C.~Dimitrakakis, R.~Coulom, and
  A.~Sumner, ``{TORCS -- Open Racing Car Simulator},''
  \url{http://torcs.sourceforge. net}, 2000, [Accessed on 2023-01-01].

\bibitem{dosovitskiy2017carla}
A.~Dosovitskiy, G.~Ros, F.~Codevilla, A.~Lopez, and V.~Koltun, ``{CARLA}: An
  open urban driving simulator,'' in \emph{Proc. of the 1st Annual Conference
  on Robot Learning}.\hskip 1em plus 0.5em minus 0.4em\relax PMLR, 2017, pp.
  1--16.

\bibitem{cleland2018dronology}
J.~Cleland-Huang, M.~Vierhauser, and S.~Bayley, ``Dronology: An incubator for
  cyber-physical system research,'' in \emph{Proc. of the 40th IEEE/ACM Int'l
  Conference on Software Engineering: New Ideas and Emerging Technologies
  Results}, 2018.

\bibitem{afzal2021gzscenic}
A.~Afzal, C.~L. Goues, and C.~S. Timperley, ``Gzscenic: Automatic scene
  generation for gazebo simulator,'' \emph{arXiv preprint arXiv:2104.08625},
  2021.

\bibitem{scanavino2019new}
M.~Scanavino, A.~Vilardi, and G.~Guglieri, ``A new facility for uav testing in
  climate-controlled environments,'' in \emph{Proc. of the 2019 Int'l
  Conference on Unmanned Aircraft Systems}.\hskip 1em plus 0.5em minus
  0.4em\relax IEEE, 2019, pp. 1436--1444.

\bibitem{grigoropoulos2020simulation}
N.~Grigoropoulos and S.~Lalis, ``Simulation and digital twin support for
  managed drone applications,'' in \emph{Proc. of the 24th Int'l Symposium on
  Distributed Simulation and Real Time Applications}.\hskip 1em plus 0.5em
  minus 0.4em\relax IEEE, 2020, pp. 1--8.

\bibitem{schmittle2018openuav}
M.~Schmittle, A.~Lukina, L.~Vacek, J.~Das, C.~P. Buskirk, S.~Rees,
  J.~Sztipanovits, R.~Grosu, and V.~Kumar, ``{OpenUAV: A UAV testbed for the
  CPS and robotics community},'' in \emph{Proc. of the 9th Int'l Conference on
  Cyber-Physical Systems}.\hskip 1em plus 0.5em minus 0.4em\relax IEEE, 2018,
  pp. 130--139.

\bibitem{schmidt2022stellauav}
T.~Schmidt and A.~Pretschner, ``{StellaUAV}: A tool for testing the safe
  behavior of uavs with scenario-based testing (tools and artifact track),'' in
  \emph{Proc. of the 2022 IEEE 33rd International Symposium on Software
  Reliability Engineering}.\hskip 1em plus 0.5em minus 0.4em\relax IEEE, 2022,
  pp. 37--48.

\bibitem{khatiri2023simulation}
S.~Khatiri, S.~Panichella, and P.~Tonella, ``Simulation-based test case
  generation for unmanned aerial vehicles in the neighborhood of real
  flights,'' in \emph{Proc. of the 16th IEEE Int'l Conference on Software
  Testing, Verification and Validation}, 2023.

\bibitem{agrawal2020model}
A.~Agrawal, J.~Cleland-Huang, and J.-P. Stegh{\"o}fer, ``Model-driven
  requirements for humans-on-the-loop multi-uav missions,'' in \emph{Proc. of
  the Tenth Int'l Model-Driven Requirements Engineering Workshop}.\hskip 1em
  plus 0.5em minus 0.4em\relax IEEE, 2020, pp. 1--10.

\bibitem{cleland2022extending}
J.~Cleland-Huang, A.~Agrawal, M.~Vierhauser, M.~Murphy, and M.~Prieto,
  ``Extending {MAPE-K} to support human-machine teaming,'' in \emph{Proc. of
  the 17th Symposium on Software Engineering for Adaptive and Self-Managing
  Systems}, 2022, pp. 120--131.

\end{thebibliography}
